\documentclass[lettersize,journal]{IEEEtran}
\usepackage{cite}
\usepackage{amsmath,amssymb,amsfonts}
\usepackage{algorithmic}
\usepackage{graphicx}
\usepackage{textcomp}
\usepackage{wrapfig}
\usepackage{subcaption}
\usepackage{booktabs}

\usepackage{multirow}
\usepackage{multicol}

\def\BibTeX{{\rm B\kern-.05em{\sc i\kern-.025em b}\kern-.08em
    T\kern-.1667em\lower.7ex\hbox{E}\kern-.125emX}}
\begin{document}
\title{Exploring FMCW Radars and Feature Maps for Activity Recognition: A Benchmark Study}

\author{Ali Samimi Fard, %\orcidlink{0009-0009-9143-4384} 
\IEEEmembership{Graduate Student Member, IEEE}, \hspace{0.2cm} Mohammadreza Mashhadigholamali, %\orcidlink{0009-0002-0811-6203} 
\IEEEmembership{Graduate Student Member, IEEE}, \hspace{0.2cm} Samaneh Zolfaghari*, %\orcidlink{0000-0003-0649-1691} 
\hspace{0.2cm} Hajar Abedi, %\orcidlink{0000-0002-4089-0373} 
\IEEEmembership{Member, IEEE}, \hspace{0.2cm} Mainak Chakraborty, %\orcidlink{0000-0003-3188-3072} 
\IEEEmembership{Senior Member, IEEE}, \hspace{0.2cm} Luigi Borzì, %\orcidlink{0000-0003-0875-6913} 
\IEEEmembership{Member, IEEE},\hspace{0.2cm}  Masoud Daneshtalab, %\orcidlink{0000-0001-6289-1521} 
\IEEEmembership{Senior Member, IEEE}, \\ George Shaker, %\orcidlink{0000-0002-1450-2138} 
\IEEEmembership{Senior Member, IEEE} 
\thanks{Ali Samimi Fard, Mohammadreza Mashhadigholamali and Luigi Borzì are with the Department of Control and Computer Engineering, Polytechnic University of Turin, Turin, Italy (\textit{Email: \{ali.samimifard, mohammadreza.mashhadigholami\}@studenti.polito.it, luigi.borzi@polito.it})}
\thanks{Samaneh Zolfaghari (Corresponding author), Mainak Chakraborty and Masoud Daneshtalab are with School of Innovation, Design and Engineering, Division of Intelligent Future Technologies, Mälardalen University, Västerås, Sweden (\textit{Email: \{samaneh.zolfaghari, mainak.chakraborty, masoud.daneshtalab\}@mdu.se})}
\thanks{Hajar Abedi and George~Shaker are with Department of Electrical and Computer Engineering, University of Waterloo, Waterloo, Ontario, Canada (\textit{Email: \{habedifi, gshaker\}@uwaterloo.ca})}}

%\IEEEtitleabstractindextext{%
% \fcolorbox{abstractbg}{abstractbg}{%
% \begin{minipage}{\textwidth}%
% \begin{wrapfigure}[11]{r}{3.16in}%
% \includegraphics[width=3.16in]{Images/abstract_plot_first_journal_new.png}%
% \end{wrapfigure}%
\maketitle

\def\journalname{This work has been submitted to the IEEE for possible publication. Copyright may be transferred without notice, after which this version may no longer be accessible.}
\markboth{\journalname}{}
\pagestyle{plain}

\begin{abstract}
Human Activity Recognition has gained significant attention due to its diverse applications, including ambient assisted living and remote sensing. Wearable sensor-based solutions often suffer from user discomfort and reliability issues, while video-based methods raise privacy concerns and perform poorly in low-light conditions or long ranges. This study introduces a Frequency-Modulated Continuous Wave radar-based framework for human activity recognition, leveraging a $60~GHz$ radar and multi-dimensional feature maps. Unlike conventional approaches that process feature maps as images, this study feeds multi-dimensional feature maps—Range-Doppler, Range-Azimuth, and Range-Elevation —as data vectors directly into the machine learning (SVM, MLP) and deep learning  (CNN, LSTM, ConvLSTM) models, preserving the spatial and temporal structures of the data. These features were extracted from a novel dataset with seven activity classes and validated using two different validation approaches. The ConvLSTM model outperformed conventional machine learning and deep learning models, achieving an accuracy of $90.51\%$ and an $F_1$-score of $87.31\%$ on cross-scene validation, and an accuracy of $89.56\%$ and an $F_1$-score of $87.15\%$ on leave-one-person-out cross-validation. The results highlight the approach's potential for scalable, non-intrusive, and privacy-preserving activity monitoring in real-world scenarios.
\end{abstract}
%\end{minipage}}}

\section{Introduction}
\IEEEPARstart{H}{uman} Activity Recognition (HAR) has become essential for applications in smart homes, healthcare monitoring, gait analysis, and fall detection~\cite{8861371, 10420485, Ullmann2023}. The rising elderly population has heightened the need for reliable activity monitoring to improve quality of life and provide timely emergency responses~\cite{10420485}. Recent advances in Machine Learning (ML), Deep Learning (DL), and low-cost sensors have made HAR technologies more accessible, enabling continuous, non-intrusive monitoring and quick interventions during emergencies~\cite{10322785,walsh2012relationship}. When it comes to sensors, they can be divided into two main categories: wearable and non-wearable~\cite{Zolfaghari2023}. 

Despite their widespread adoption, wearable sensors come with several challenges. They can cause discomfort for users, rely heavily on battery life, and often become less reliable during activities like bathing or sleeping~\cite{Zhao2022}. Their accuracy also depends on where they are placed on the body~\cite{Li2018}. These issues highlight the need for alternative solutions. In this context, non-wearable sensors offer a different approach by gathering data from the surrounding environment without needing users to carry them~\cite{9807403, Zhang2017,10.1145/3310194,Ullmann2023}. However, they come with their own set of challenges, such as privacy concerns with video cameras~\cite{9727166}, and limitations from environmental factors (e.g., sunlight, obstructions) for infrared sensors~\cite{Ullmann2023, Seifert2021,Zolfaghari2023}.

In recent years, radar-based sensing, particularly Frequency-Modulated Continuous Wave (FMCW) radar, has become a promising solution for non-intrusive monitoring without wearable devices or cameras~\cite{Ullmann2023,miazek2024human}. Known for high accuracy and robustness, even in low-light and obstructed environments~\cite{app132312728, miazek2024human,9727166}, FMCW radars transmit chirp signals and use Doppler shifts to capture detailed motion information. This includes larger movements like walking and sitting, as well as precise actions like picking up objects or falling~\cite{Ullmann2023, electronics12020308}.

Notably, while HAR has made great strides with diverse sensor technologies, there is still a need for accurate, non-intrusive, and privacy-preserving solutions. This work contributes a new benchmark dataset for HAR, collected in realistic environments with a variety of activities, supporting robust model development and evaluation. Additionally, a novel data processing approach is presented that treats multi-dimensional radar feature maps as data vectors, preserving the spatiotemporal characteristics. Together, these contributions aim to enhance the performance and robustness of the proposed approach. This study introduces an FMCW radar-based framework for HAR, with the following key contributions:

\begin{itemize}
    \item We collected a dataset in a realistic setting using a low-resolution $60\,\text{GHz}$ mmWave FMCW radar, covering intricate and less-studied activities for real-life representation. It will be publicly available upon publication to support further HAR research with mmWave FMCW radar.
    \item We have structured RD, RA, and RE feature maps to comprehensively capture motion patterns and spatial relationships, treating them as 3D data vectors instead of 2D images.
    \item We have shown that training an appropriate DL model with data from just three subjects is sufficient to achieve reliable and generalizable results across a range of human activities, provided that there are fixed objects in the environment.
    \item We have validated the performance of the proposed approach using two distinct cross-validation methods on various ML and DL models.
\end{itemize}

The paper is structured as follows: Section~\ref{related_work} reviews related work. Section~\ref{methodology} details the methodology, radar setup, and data processing. Section~\ref{eval} presents experiments, results, and performance comparisons. Section~\ref{discussion} highlights key contributions and applications. Finally, Section~\ref{conclusion} concludes and suggests future research.

\section{Related work}
\label{related_work}

Recent advancements in HAR systems have leveraged sophisticated DL models and innovative feature extraction techniques to enhance accuracy and efficiency~\cite{Abedi2023}.  

For instance, Abdu et al.~\cite{9727166} used micro-Doppler spectrogram images from the University of Glasgow dataset~\cite{fioranelli2019radar} and employed a transfer learning approach to compare three different pre-trained models and their combinations. They also applied a channel attention module to improve performance in classifying six activities. Their experiments achieved $96.86\%$ accuracy with the VGG-19 network and $99.71\%$ accuracy using a combination of AlexNet and VGG-19, enhanced by the Canonical Correlation Analysis (CCA) feature fusion method.
Kim et al.~\cite{9743913} combined Range-Time-Doppler (RTD) maps with a Range-Distributed Convolutional Neural Network (RD-CNN), achieving $96.49\%$ accuracy in familiar environments using the University of Glasgow dataset~\cite{fioranelli2019radar}.
Hsu et al.~\cite{10209420} developed a fall detection system using range-Doppler images and a CNN model to detect sudden speed changes, followed by a bidirectional long short-term memory network (Bi-LSTM) to confirm falls, achieving nearly $96\%$ accuracy.
Ding et al.~\cite{9881387} used dynamic range-Doppler frames (DRDF) and ST-ConvLSTM networks to capture and integrate spatial-temporal features, achieving $96.5\%$ accuracy in classifying six motions performed by 16 subjects.
Bhavanasi et al.~\cite{bhavanasi2022patient} evaluated various ML algorithms on Micro-Doppler and Range-Doppler maps from radar data in hospital environments. CNN models outperformed others, achieving high accuracy in classifying ten activities performed by $29$ subjects.

Despite significant progress, most radar-based HAR systems use image-based feature maps~\cite{9727166, Zhao2022, 9743913, 10209420}. These have limitations like poor interpretability due to high similarity between activity images and susceptibility to noise and artifacts~\cite{10013652}.
Furthermore, training DL models on image-based feature maps often requires large labeled datasets, which can be challenging to obtain~\cite{9743913}. To address this, we propose an alternative approach that directly uses multi-dimensional radar feature maps—Range-Doppler (RD), Range-Azimuth (RA), and Range-Elevation (RE)—
%in a Convolutional Long Short-Term Memory (ConvLSTM) network 
as 3D data vectors. %This preserves spatial and temporal characteristics, improving the model's ability to capture complex activity patterns. 
Validation across various conventional and modern models demonstrated robust performance with data from just three participants, minimizing the need for large datasets. Additionally, real-world data collection poses challenges, as many studies use controlled lab environments that do not reflect real-life conditions~\cite{9881387}.To advance HAR with new radar systems and benchmark datasets, this study introduces a novel dataset from a low-resolution $60\,\text{GHz}$ mmWave FMCW radar and a framework for its pre- and post-processing in HAR applications.

\section{Methodology}
\label{methodology}
\begin{figure*}[htbp]
    \centering
    \includegraphics[width=\textwidth]{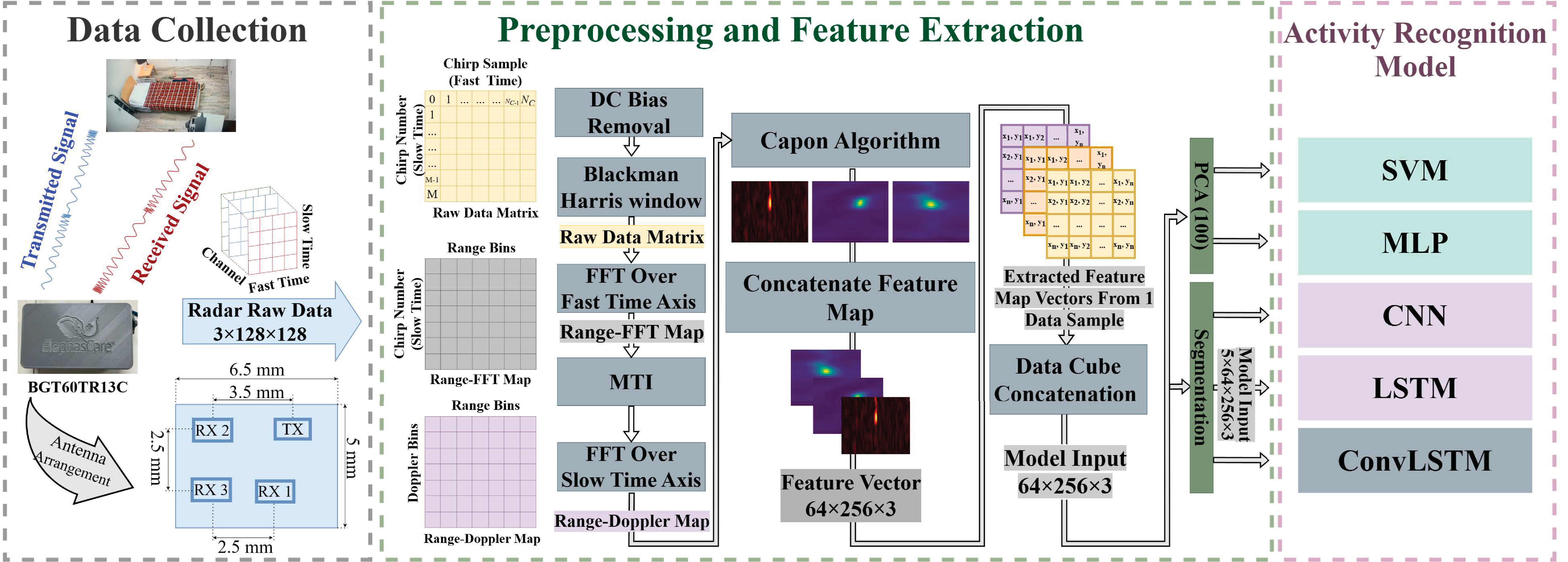}
    \caption{Overview of the proposed framework for FMCW radar-based HAR.}
    \label{flow chart}
\end{figure*}

This section provides an overview of the functional prototype of our proposed framework, which is depicted in Fig.~\ref{flow chart}.
Each component is described in detail in the following subsections.

\subsection{Radar Setup and Data Collection}
\label{radar_setup_data_collection}
This study uses the BGT60TR13C, an FMCW radar system from Infineon Technologies AG~\cite{Infineon:BGT60TR13C}. It has a single transmitter and three receivers, operating in the 58-63.5 GHz band with a configurable chirp duration. The antennas are in an L-shaped configuration, with RX1 and RX3 for azimuth and RX2 and RX3 for elevation angle measurements~\cite{site2}. Table~\ref{Radar Configuration and Specification} lists the radar configuration parameters used in this study.
%Check again with Hajar

\begin{table}[htbp]
    \centering
    \caption{Radar configuration and specification.}
    \label{Radar Configuration and Specification}
    \begin{tabular}{c c}
    \hline
        \textbf{Parameters} & \textbf{Value} \\ \hline \hline
         Radar Model& BGT60TR13C  \\
         Start Frequency & 61 GHz  \\
         End Frequency & 62 GHz  \\
         Transmit Output Power & 5 dBm\\
         ADC Sampling Rate & 2 Msps \\
         Frame Rate & 10 \\
         Chirps Per Frame & 128 \\
         Number of Tx Antennas & 1  \\
         Number of Rx Antennas & 3 \\
         Range resolution & 15 cm \\
         Max Unambiguous Range & 4.8 m \\
         \hline
    \end{tabular}
\end{table}

To collect data, the radar-based sensor is arranged to ensure optimal bedroom coverage, minimizing side-lobe interference and maximizing performance. It is mounted at a height of $210$~cm and angled downward at $30$ degrees. Three healthy subjects participated in the study, performing activities over $16$ trials. Two subjects completed six trials each, and the third completed four. Each trial began with a 1-minute data collection in an empty room. Subjects then performed: walking for $2$ minutes, sitting on a bed for $2$ minutes, lying on the bed for $5$ minutes, and lying on the floor for $5$ minutes. In some sessions, they also sat on a chair for $2$ minutes. Data was collected continuously, capturing both steady-state activities and transitions. Transitions were grouped into a single ``Transition'' class to address classification challenges and balance the dataset. The data was categorized into seven activity classes: Empty Room, Walking, Sitting on the Bed, Sitting on a Chair, Lying on the Bed, Lying on the Floor, and Transition. The data collection protocol was approved by the University of Waterloo's Research Ethics Committee. All activities complied with applicable safety and ethical standards. The dataset will be made available at GitHub\footnote{\underline{https://github.com/}} upon acceptance of this manuscript.

\subsection{Data Preprocessing and Feature Extraction}
\label{data_preprocessing_feature_extraction}

In FMCW radar systems, chirp sequences are transmitted via the \(TX\) antenna, with reflections captured by \(RX\) antennas. The received data forms a three-dimensional array of dimensions \(C \times N \times M\), where \(C\) represents the number of channels, \(N\) denotes the number of chirps per frame, and \(M\) indicates the number of samples per chirp. The data structure comprises \textit{fast time} rows (single chirp/range bin data) and \textit{slow time} columns (same-sample data across chirps)~\cite{Peng2019}.

\subsubsection{Blackman-Harris Window}

The preprocessing pipeline involves DC bias removal to eliminate low-frequency noise and artifacts~\cite{9136197}. Subsequently, a Blackman-Harris window function is applied to mitigate spectral leakage in the frequency domain. This windowing technique gradually attenuates signal amplitudes at the boundaries, reducing abrupt transitions that could introduce spurious frequency components during Fourier transformation. This approach enhances the fidelity of spectral analysis, particularly when employing the Fast Fourier Transform (FFT).

\subsubsection{Range-FFT Map}
Range detection is performed by computing the FFT along the fast-time axis, where spectral peaks correspond to target distances~\cite{zeng2022walking}. The resulting Range-FFT map provides a frequency-domain representation of target reflections. The radar transmits consecutive chirps separated by a fixed time interval to estimate target velocity. Each reflected chirp undergoes a Range-FFT to determine the target position. While the peaks in the Range-FFT spectrum align for both chirps, their phases differ due to target movement. This phase shift provides information about the velocity of the target~\cite{lovescu2020fundamentals, Abedi2021}.

\subsubsection{Moving Target Indicator (MTI)}
The signal consists of two main types of reflections. The first is clutter, which refers to echoes from stationary objects in the environment. The second type originates from moving objects, particularly individuals engaged in daily activities. A clutter removal algorithm is employed to reduce the impact of clutter.

\begin{figure*}[!ht]
    \centering
    \includegraphics[width=0.9\linewidth]{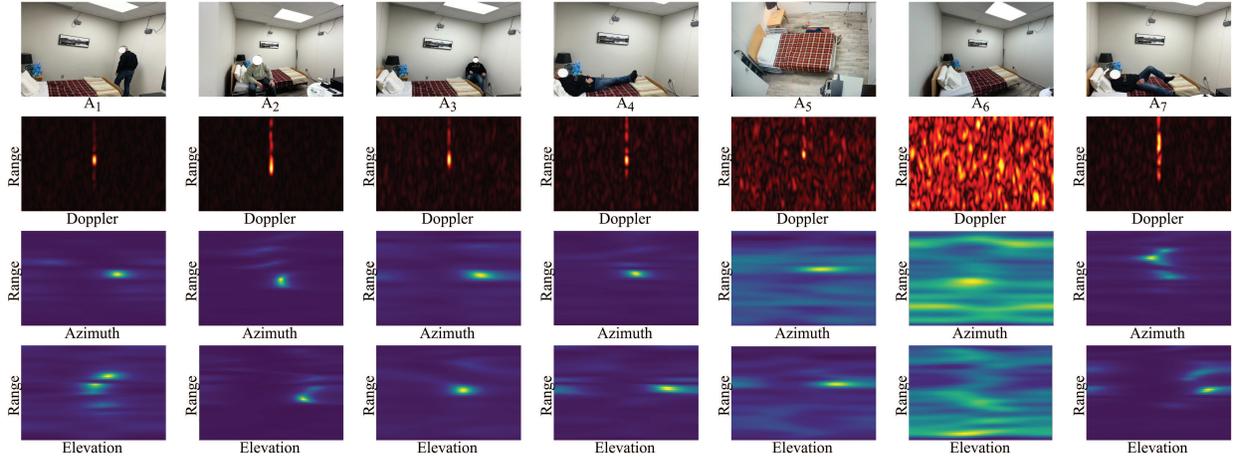}
    \caption{Examples of a participant performing the activities with corresponding feature maps. ($A_1$) walking, ($A_2$) sitting on the bed, ($A_3$) sitting on the chair, ($A_4$) lying down on the bed, ($A_5$) lying down on the floor, ($A_6$) empty room, ($A_7$) transition.}
    \label{activities_plots}
\end{figure*} 

The MTI implements linear filtering to suppress clutter while preserving dynamic target signatures. In FMCW radar systems, the Finite Impulse Response (FIR) implementation offers a good balance of simplicity and effectiveness~\cite{8331965}. At each time step, the maximum absolute value across the slow time dimension for each range bin is denoted as \(r_{i,max}\). The MTI filter output \(t_i\) is then calculated as a weighted average of this peak value and the previous filter output \(t_{i-1}\), using a weighting factor \(\alpha\):

\begin{equation}
    t_i = \alpha \cdot r_{i,max} + (1 - \alpha) \cdot t_{i-1}
\end{equation}

At the initial time step (\(t_0\)), the filter output \(t_i\) is initialized to zero. For each range bin, the MTI filter removes the influence of stationary objects by subtracting \(t_{i-1}\) from \(r_{i,max}\), resulting in the filtered FFT value \(r_{i,filt}\) \cite{will2019human}:

\begin{equation}
    r_{i,filt} = \left| r_{i,max} -  t_{i-1} \right|
\end{equation}

This method of subtracting an estimate of stationary background clutter effectively eliminates static targets while having minimal effect on slow-moving objects. FIR MTI filters are preferred for their simple design, adjustable parameters, and linear phase response~\cite{8331965, will2019human}.

\subsubsection{Range-Doppler Map}
The processing sequence continues with a second FFT applied along the vertical axis to extract Doppler information for each channel. The output is the Range-Doppler map (RDM).

\subsubsection{Capon Algorithm}
Three-dimensional object localization requires precise determination of range, as well as vertical and horizontal angles. The Capon algorithm, also known as the MVDR beamformer, is applied to the RD map to enable accurate position estimation and support the generation of RA and RE maps. It is designed to estimate the Angle of Arrival (AoA) in noisy and interference-prone environments~\cite{1449208, alizadeh2019low}. The algorithm minimizes the output power of an array while maintaining a distortionless response at the desired angle, effectively enhancing the signal-to-interference-plus-noise ratio by suppressing unwanted signals and focusing on the target direction. This process is achieved through constrained optimization.
For more details on the Capon beamforming algorithm, refer to our previous work~\cite{abedi2021ai}.

The final 3D data structure, called the data cube, is formed by combining the RD, RA, and RE feature maps. Before feeding data to DL models, segmentation is done on sets of five consecutive data cubes to balance granularity and efficiency. If all five data cubes have the same label activity, they are merged into a single segment. If activities differ, such as three matching and two different, the matching cubes are excluded, and segmentation pauses. It resumes with the next group of data cubes for the subsequent activity, ensuring consistency and preventing conflicting labels from merging. 

Figure~\ref{activities_plots} illustrates examples of a participant performing various activities, along with the corresponding feature maps. As shown in the figure, when the subject is moving, there is significantly less clutter and noise, allowing us to clearly observe the moving subject (Fig.\ref{activities_plots} $A_1$ to $A_5$). This clarity is achieved through our implementation of the MTI. In contrast, the feature map for the `Empty room' (Fig.\ref{activities_plots} $A_6$) shows visible clutter and noise, which are echoes from stationary objects.

\subsection{Activity Recognition Models}
\label{activity_recognition_model}

In this study, we evaluated the performance of two conventional ML classifiers—Support Vector Machines (SVM) and Multi-Layer Perceptrons (MLP)—for recognizing various activities. To address the limitations of ML models, we further implemented LSTM, CNN and ConvLSTM models. The configurations of these models are outlined below, with unspecified parameters set to their default values:

\begin{itemize}
    \item \textbf{SVM~\cite{cortes1995support}:} The kernel was set to a radial basis function (RBF) with regularization term $C=10$ and probability=True.
    \item \textbf{MLP~\cite{rumelhart1986learning}:} The network consisted of two hidden layers with 128 and 64 neurons, and Rectified Linear Unit (ReLU) activation function. Training was performed using the Adaptive moment estimation (Adam) optimizer with a learning rate of $1\times10^{-3}$, a maximum of 300 iterations, and random\_state=$42$.
    \item \textbf{CNN~\cite{bhavanasi2022patient}:} The model consists of four 3D convolutional blocks with 8, 16, 32, and 64 filters respectively, each using a kernel size of ($3 \times 3 \times 3$) and Exponential Linear Unit (ELU) activation function. Each block is followed by a $1 \times 2 \times 2$ max pooling layer. The network includes two fully connected layers: one with 128 neurons and another with 7, 6, 5, or 4 neurons (corresponding to the number of activities being classified). Dropout is applied after each layer at progressively increasing rates from 0.2 to 0.5, excluding the output layer. The network is trained using the Adam optimizer with a cross-entropy loss function, a learning rate of $1 \times 10^{-3}$, and a maximum of 100 epochs with early stopping (patience = 10 epochs).
    
    \item \textbf{LSTM~\cite{8861371}:} The architecture comprised two Bi-LSTM layers with 256 units and a dropout of 0.5. The network includes two fully connected layers: one with 128 neurons and another with 7, 6, 5, or 4 neurons (corresponding to the number of activities being classified). The model was trained using the Adam optimizer with a decay factor of 0.9 and an initial learning rate of $1 \times 10^{-3}$. This learning rate was reduced to 10\% of its initial value at the 200th epoch, with training continuing for a maximum of 400 epochs with early stopping (patience = 40 epochs).
    \item \textbf{ConvLSTM~\cite{shi2015convolutional}:} The architecture consists of one ConvLSTM block with 32 filters, kernel size of ($3 \times 3$), and ReLU activation function. This is followed by batch normalization, 3D MaxPooling ($1 \times 2 \times 2$), and dropout (0.3). The output is connected to a fully connected layer with 64 neurons and ReLU activation, followed by dropout (0.5). The final fully connected layer contains 7, 6, 5, or 4 neurons (corresponding to the number of activities being classified) with softmax activation. The model was trained using Stochastic Gradient Descent (SGD) with momentum 0.9 and weight decay $1 \times 10^{-4}$, optimizing the categorical cross-entropy loss function. Training utilized a learning rate of $1 \times 10^{-4}$, batch size of 64, and a maximum of 100 epochs with early stopping (patience = 10 epochs).
    \end{itemize}

These hyperparameters were determined via a systematic grid search to balance computational efficiency and model performance.
%All ML and DL models were trained and evaluated using the same training, validation, and test splits of the first strategy, which will be described in the Section~\ref{eval}. 
The models employed preprocessed RD, RA, and RE data from the FMCW radar as input features. Also, to accelerate training and reduce overfitting, we applied principal component analysis (PCA) with 100 components before feeding the data to the ML models. Our framework processes the spatiotemporal characteristics of FMCW radar outputs through three-dimensional (3D) input tensors. 
\section{Experimental Evaluation}
\label{eval}
This section presents experimental results obtained from an FMCW radar dataset, including analyses of feature maps, activities, and ML and DL models. We detail the setup, results, and performance analysis.

\subsection{Experimental Setup}
To validate our approach, we employed two distinct strategies. In the first strategy, Cross-Scene Validation (CSV), we used $80\%$ of the data from 14 of the 16 distinct scenes for training and reserved the remaining $20\%$ for validation. The two remaining scenes were held out for testing.
In the second strategy, Leave-One-Person-Out Cross-Validation (LOPO-CV), we repeated the procedure three times—each time leaving one subject out—and computed the accuracy and $F_1$-score. Specifically, in each fold, we trained the model using $80\%$ of the data from the two subjects, reserved the remaining $20\%$ for validation, and used the unseen subject’s data for testing. For example, in the first iteration, the data of the first subject was held out for testing, while data from the second and third subjects was split for training and validation; in the subsequent iterations, the roles were rotated accordingly. Tables~\ref{activity_table1} and \ref{leave_person_out_data_distribution} summarize the data distribution for each validation strategy.

\begin{table}[htbp]
\centering
\huge
\caption{Activity sample counts under CSV approach.}
\label{activity_table1}
\resizebox{\columnwidth}{!}{%
\begin{tabular}{lrrrrr}
\toprule
\textbf{Label}         & \textbf{Training} & \textbf{Validation} & \textbf{Testing} & \textbf{Total} & \textbf{Symbol}   \\
\midrule
Walking  & 6,495  & 1,625  & 1,250  & 9,370   & \(A_1\)    \\
Sitting on the Bed             & 5,130  & 1,285  & 705  & 7,120   & \(A_2\)     \\
Sitting on the Chair           & 1,255  & 310  & 1,205  & 2,770   & \(A_3\)      \\
Lying Down on the Bed          & 12,870 & 3,220  & 1,540  & 17,630  & \(A_4\)      \\
Lying Down on the Floor        & 9,810 & 2,455 & 3,040  & 15,305  & \(A_5\)       \\
Empty Room                     & 3,405  & 850  & 640  & 4,895   & \(A_6\)   \\
Transition                     & 4,305  & 1,075 & 790 & 6,170   & \(A_7\)       \\
\midrule
\textbf{Total}                 & 43,270 & 10,820  & 9,170  & 63,260 \\
\bottomrule
\end{tabular}
}
\end{table}

\begin{table}[!htbp]
\centering
\footnotesize
\vspace{0.1cm}
\caption{Activity sample counts per subject in LOPO-CV approach. ($A_1$) walking, ($A_2$) sitting on the bed, ($A_3$) sitting on the chair, ($A_4$) lying down on the bed, ($A_5$) lying down on the floor, ($A_6$) empty room, ($A_7$) transition. ($S_1$) subject one, ($S_2$) subject two and ($S_3$) subject three.}
\label{leave_person_out_data_distribution}
\begin{tabular}{lccc}
\toprule
\textbf{Label} & \textbf{$S_1$} & \textbf{$S_2$} & \textbf{$S_3$} \\
\midrule
\(A_1\)            & 3020         & 3800         & 2583         \\
\(A_2\)     & 1800         & 2661         & 2684         \\
\(A_3\)   & 1205         & 1568         & -            \\
\(A_4\)       & 5327         & 6142         & 6181         \\
\(A_5\)     & 5993         & 6252         & 3082         \\
\(A_6\)         & 1514         & 2030         & 1382         \\
\(A_7\)         & 1749         & 2170         & 2298         \\
\bottomrule
\end{tabular}
\end{table}

To evaluate the models, we developed a normalization technique. Mean and standard deviation were computed from each feature channel using the combined training and validation data. These parameters normalized both the training-validation subset and the test set, ensuring uniform scaling and improving the learning process. All computations were performed on a high-performance system featuring an AMD Ryzen  {\(7~6800H\)} processor, \textit{32}~GB of system memory, and an NVIDIA GeForce RTX \textit{3060} GPU running on Windows~\textit{11}. For model development, we used Python~\textit{3.8.19}, Scikit-learn~\textit{1.3.2}, and TensorFlow~\textit{2.10}.

For comprehensive models evaluation, we employed different metrics for each validation approach. In the CSV strategy, we assessed each model performance through accuracy, precision, recall, and $F_1$-score. For the LOPO-CV approach, we calculated accuracy and $F_1$-score during classification of different numbers of activity of each subject, ultimately computing the average performance across all three subjects.

\subsection{Experimental Results}
%Conventional ML models struggle to distinguish RD, RA, and RE maps, making subtle pattern extraction challenging. DL models like CNNs excel at feature extraction, while ConvLSTM enhances radar data processing~\cite{s20041105, KONG2022110743}. Traditional LSTMs model temporal sequences but lack spatial awareness, whereas CNNs capture spatial features but not temporal dependencies. A simple CNN-LSTM combination may miss both aspects~\cite{9881387}. ConvLSTM overcomes this by integrating convolution into LSTM gates, enabling joint spatial-temporal modeling. This makes ConvLSTM particularly suited for joint spatial-temporal analysis in human activity recognition~\cite{9881387,majd2019motion}.

\subsubsection{CSV}

%This study compares the performance of ML and DL models for HAR using RD+RA+RE features from an FMCW radar system. 
%We evaluated the ConvLSTM model to assess its performance when using all three feature maps in combination and to evaluate the impact of using each feature map individually or in pairwise combinations. 
The performance comparison of different models evaluated over multiple activity categories with combination of RD, RA, and RE (RD+RA+RE) feature maps as models' input is presented in Table~\ref{model_performance_comparison}.

Table~\ref{model_performance_comparison} illustrates a consistent performance pattern across all activity sets: DL models (CNN, LSTM, and ConvLSTM) significantly outperform traditional ML approaches (SVM and MLP). In particular, ConvLSTM consistently achieves the highest performance across all activity sets, with peak accuracies of $90.51\%$ for 7-activity classification and $97.87\%$ for 4-activity classification. %The performance advantage of ConvLSTM over other models remains consistent. 
This trend indicates that ConvLSTM's ability to capture both spatial and temporal features provides a substantial advantage for HAR tasks using radar data.

\begin{figure*}[!ht]
\centering
\begin{subfigure}[!b]{0.43\columnwidth}
        \centering
    \includegraphics[width=\columnwidth]{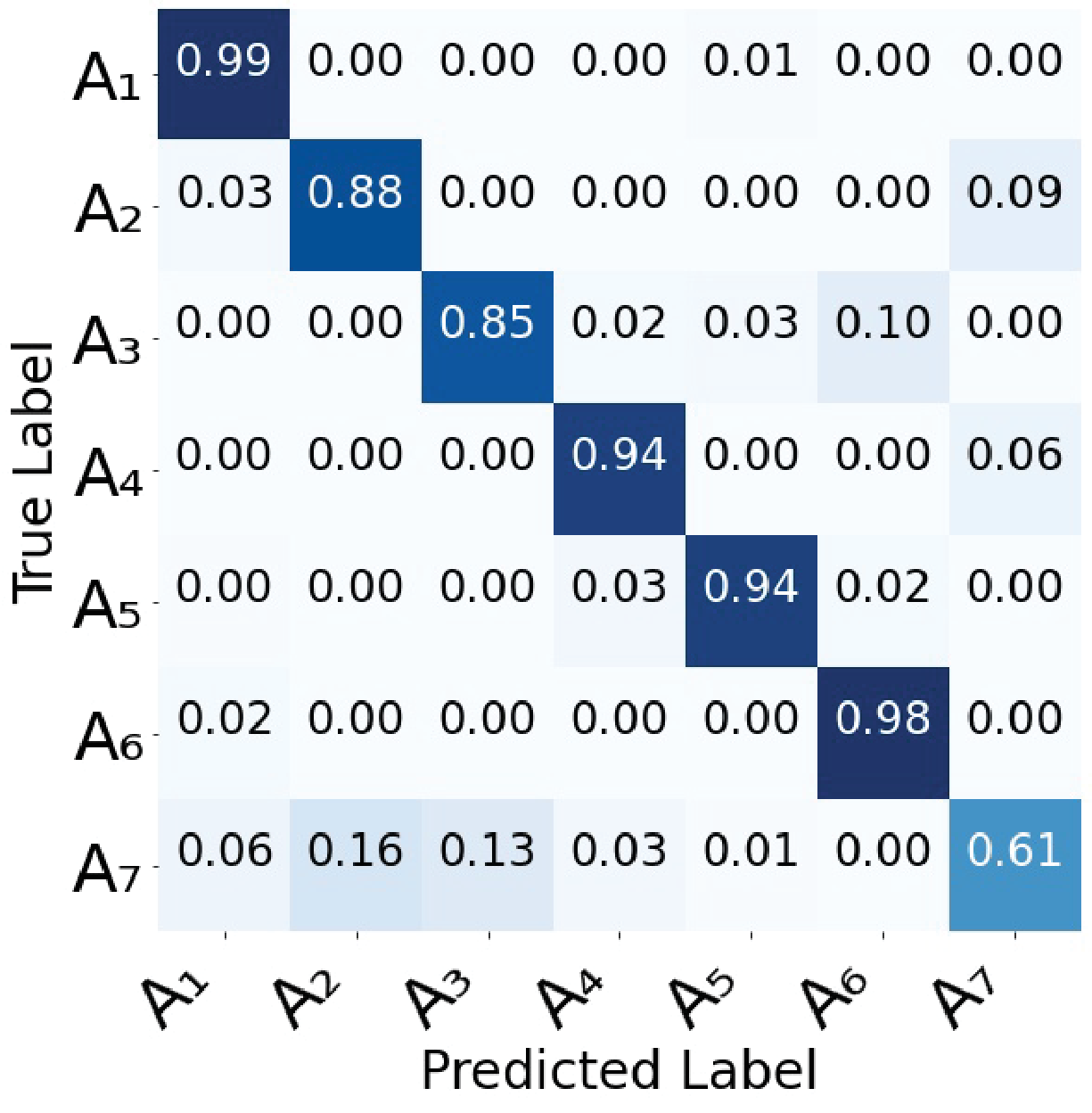}
         \caption{7 activities.}
         \label{7_cm}
     \end{subfigure}
    \centering
    \begin{subfigure}[!b]{0.43\columnwidth}
        \centering
     \includegraphics[width=\columnwidth]{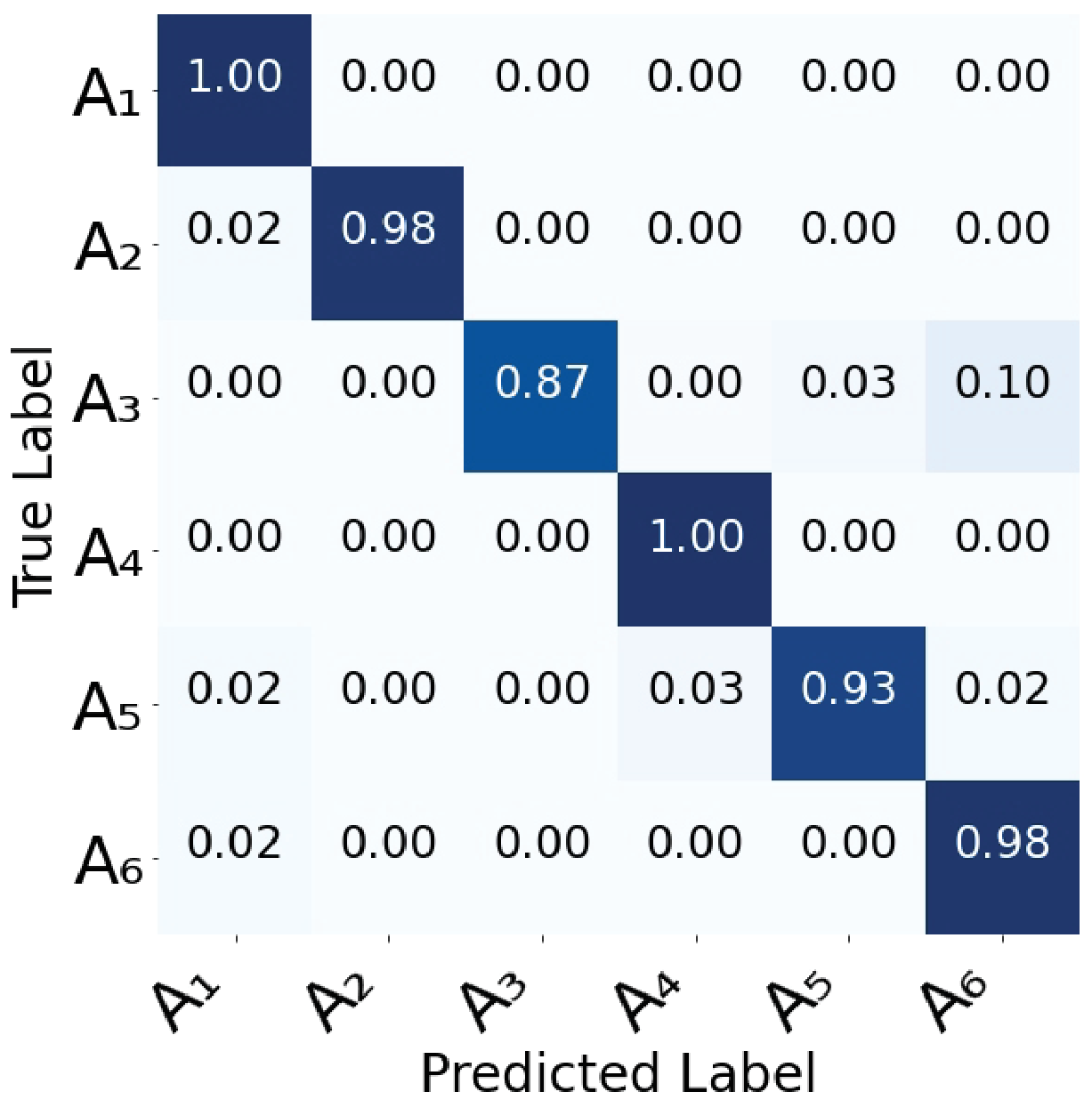}
         \caption{6 activities.}
         \label{6_cm}
     \end{subfigure}
    \begin{subfigure}[!ht]{0.43\columnwidth}
        \centering
         \includegraphics[width=\columnwidth]{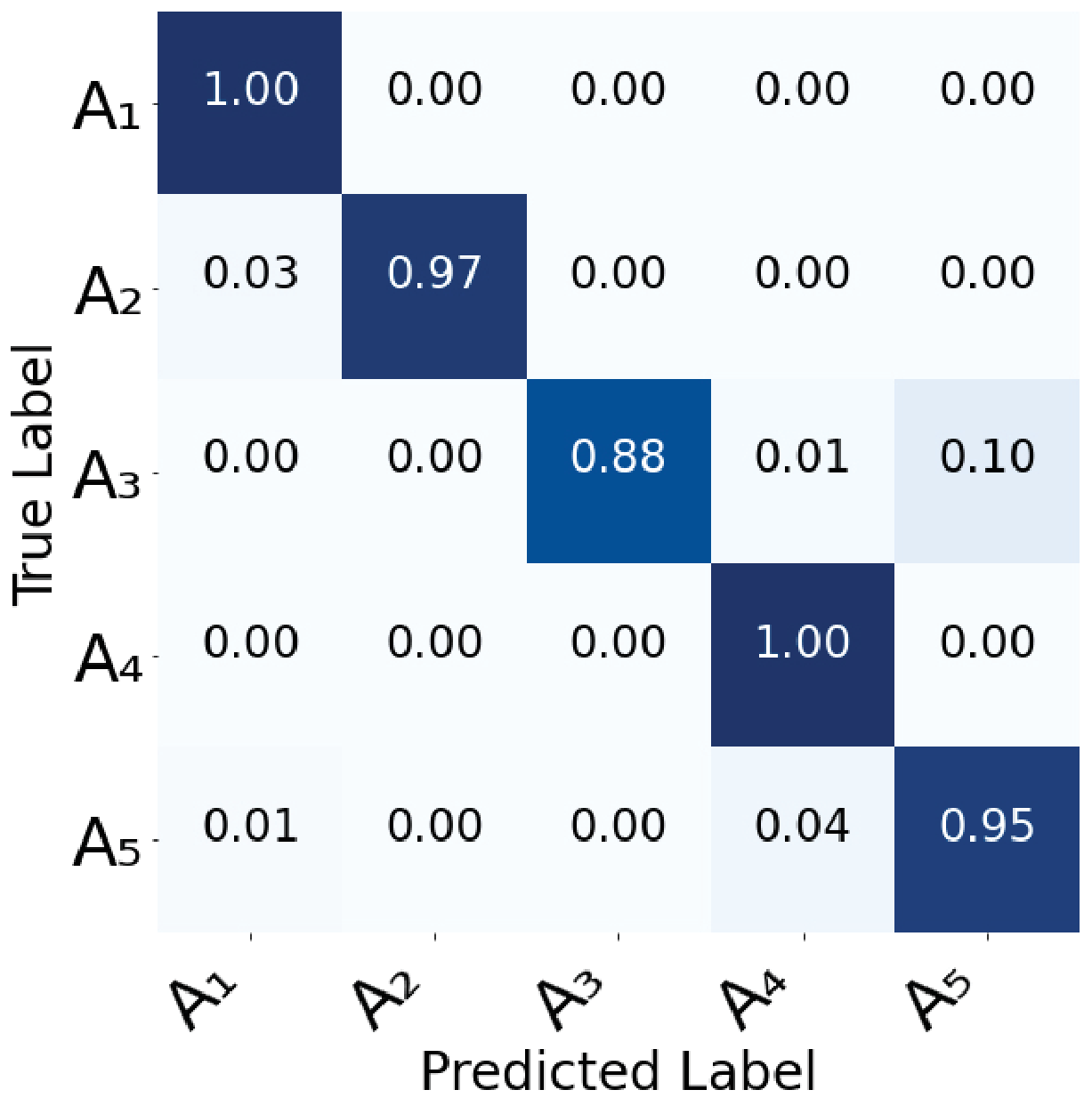}
         \caption{5  activities.}
         \label{5_cm}
     \end{subfigure}
    \begin{subfigure}[!b]{0.43\columnwidth}
        \centering
     \includegraphics[width=\columnwidth]{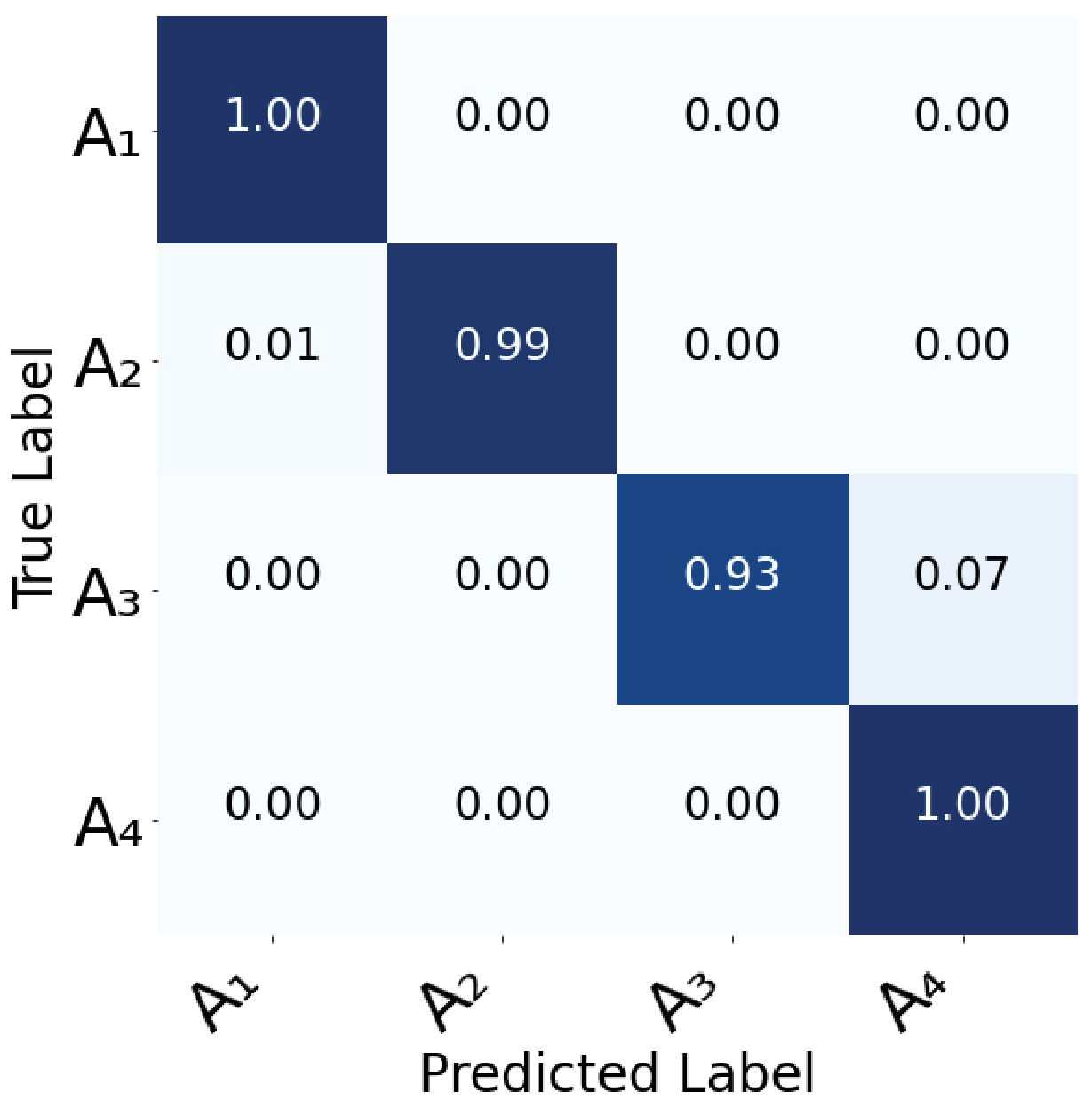}
         \caption{4 activities.}
         \label{4_cm}
     \end{subfigure}
    \caption{Confusion matrices for the ConvLSTM model on various activity sets under CSV (RD+RA+RE inputs).}
    \label{convlstm_cm}
\end{figure*}

\begin{figure*}[!ht]
\centering
\begin{subfigure}[!b]{0.43\columnwidth}
        \centering
    \includegraphics[width=\columnwidth]{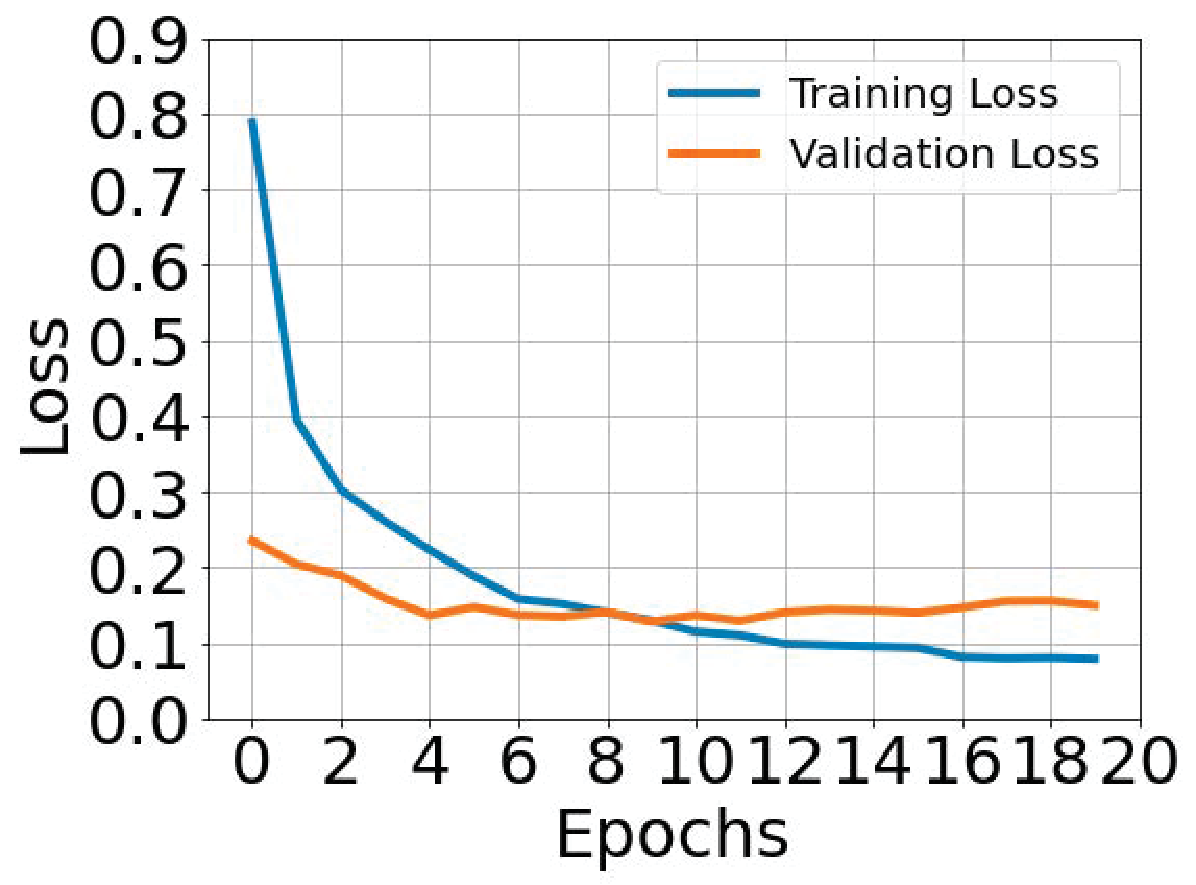}
         \caption{7 activities.}
         \label{7_loss}
     \end{subfigure}
    \centering
    \begin{subfigure}[!b]{0.43\columnwidth}
        \centering
     \includegraphics[width=\columnwidth]{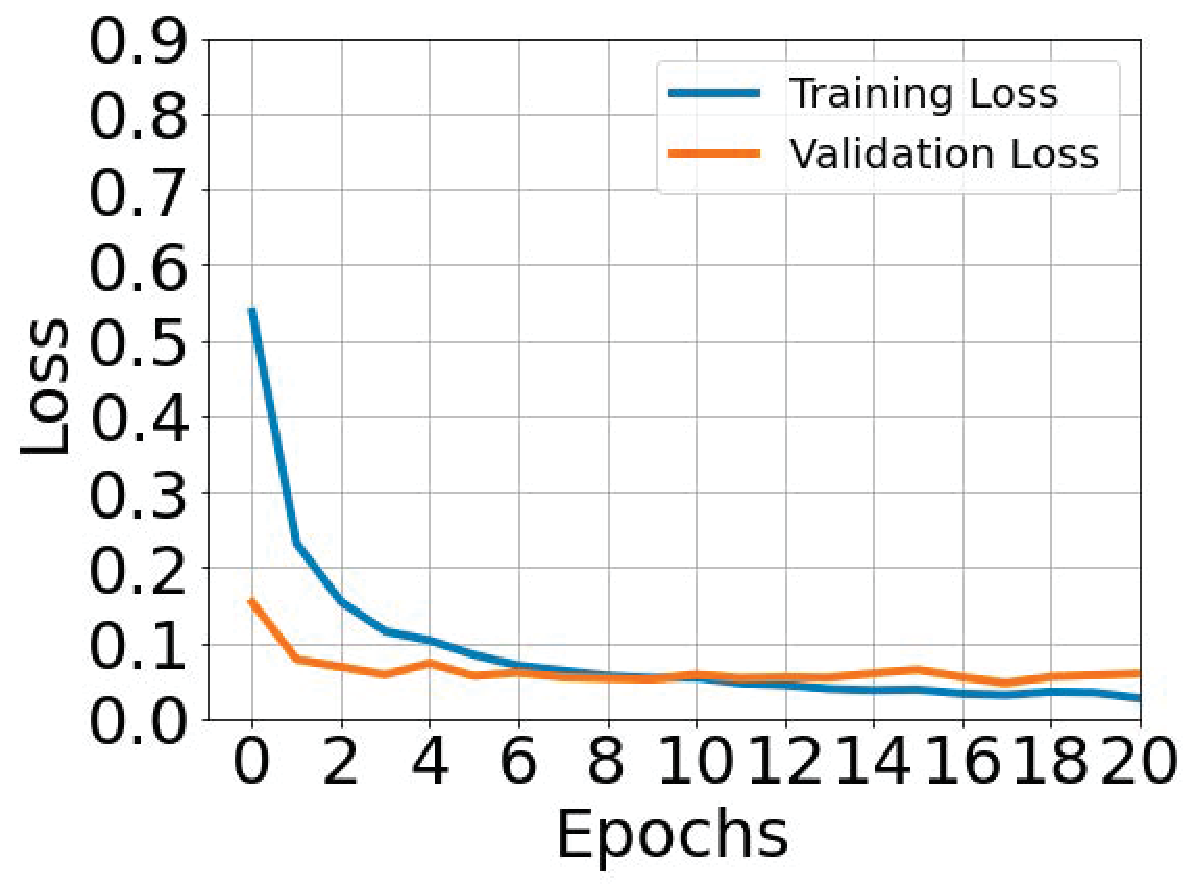}
         \caption{6 activities.}
         \label{6_loss}
     \end{subfigure}
    \begin{subfigure}[!ht]{0.43\columnwidth}
        \centering
         \includegraphics[width=\columnwidth]{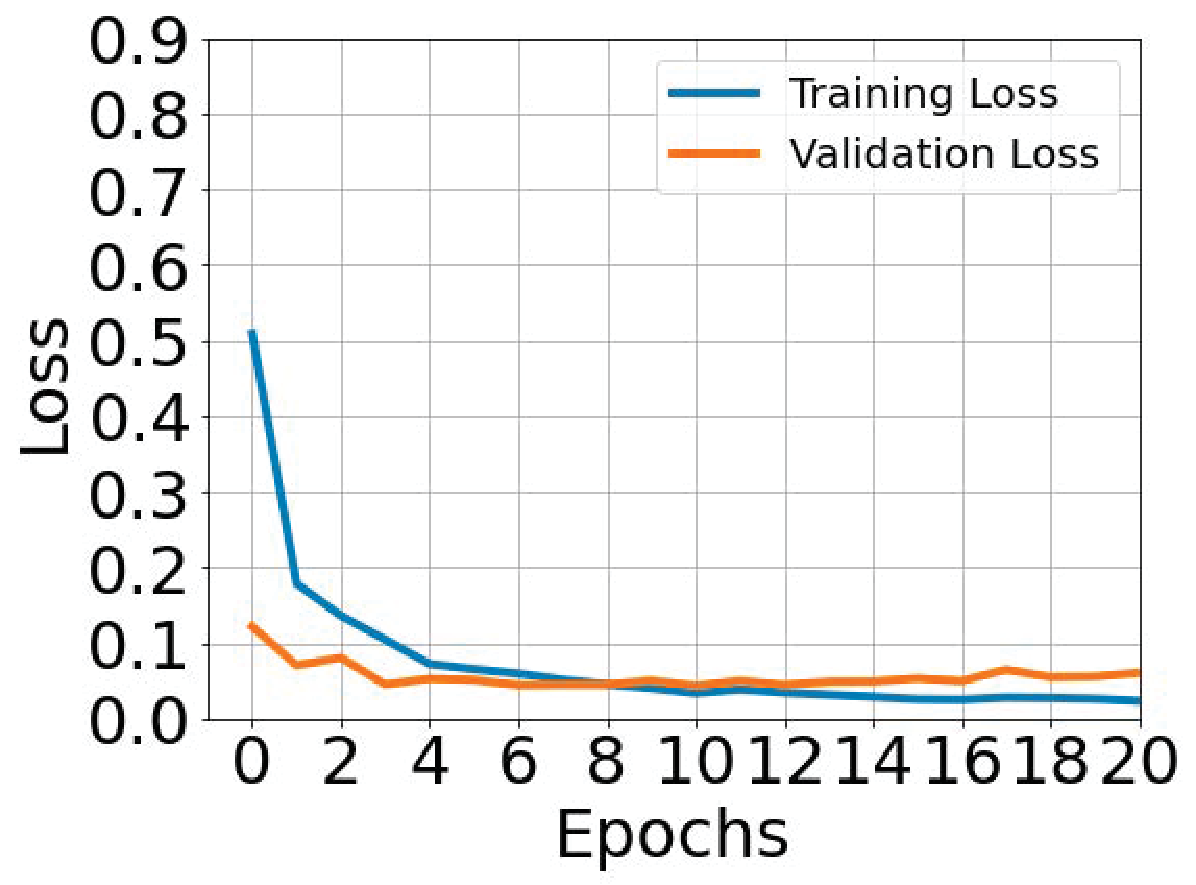}
         \caption{5 activities.}
         \label{5_loss}
     \end{subfigure}
    \begin{subfigure}[!b]{0.43\columnwidth}
        \centering
     \includegraphics[width=\columnwidth]{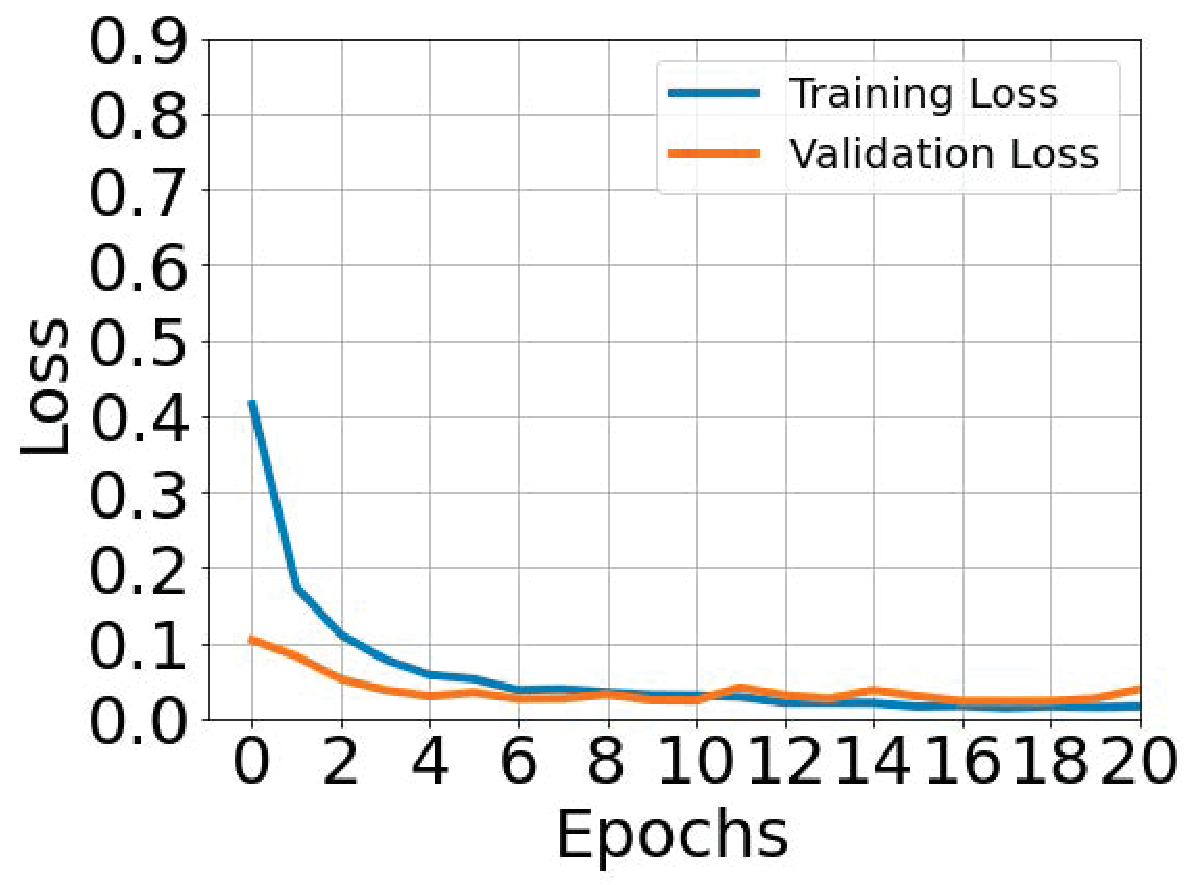}
         \caption{4 activities.}
         \label{4_loss}
     \end{subfigure}
    \caption{Training and validation loss curves for the ConvLSTM model under CSV (RD+RA+RE inputs).}
    \label{convlstm_ls}
\end{figure*}

The classification results of the ConvLSTM model with different feature inputs over various numbers of activities are summarized in Tables~\ref{model_performance_comparison1}--\ref{model_performance_comparison4}. Additionally, Fig.~\ref{convlstm_cm} shows the confusion matrices for different tasks, while Fig.~\ref{convlstm_ls} illustrates the training and validation loss curves of the ConvLSTM model across different activity sets using the RD+RA+RE feature maps.

As shown in Tables~\ref{model_performance_comparison1}--\ref{model_performance_comparison4}, classification performance improves as the number of classes decreases. Specifically, the $F_1$-score increases from $87.31\%$ for seven classes to $98.05\%$ for four classes. The RE map consistently achieves the best results across all tasks among single-feature inputs. For pairwise combinations, except for four classes, RD+RE maps outperform other combinations. Integrating all three features yields the highest performance, slightly surpassing RD+RE maps. This highlights the significance of combining motion (Doppler) and spatial dimensions (azimuth and elevation) to comprehensively represent activities, thereby enhancing classification accuracy.

The training process, shown in Fig.~\ref{convlstm_ls}, demonstrates a steady decrease in both training and validation loss, reaching suitably low levels. This reflects effective model learning, the absence of overfitting, and strong generalization capability.

\begin{table}[!htbp]
\scriptsize
\centering
\vspace{0.1cm}
\caption{Performance metrics for various models under CSV with RD+RA+RE inputs.}
\label{model_performance_comparison}
\begin{tabular}{c|c|c|c|c|c}
\toprule
\multirow{2.3}{*}{\textbf{\#Activities}} 
  & \multirow{2.3}{*}{\textbf{Model}} 
  & \multicolumn{4}{c}{\textbf{Metrics (\%)}} \\ 
\cmidrule{3-6}
 &  & \textbf{Accuracy} & \textbf{Precision} & \textbf{Recall} & \textbf{\(F_1\)-score} \\ 
\midrule

%------------------------ 7 Activities ------------------------
\multirow{7}{*}{7} 
 & SVM & 70.97 & 72.20 & 74.17 & 70.41 \\ 
 \cmidrule{2-6}
 & MLP & 68.71 & 66.35 & 68.76 & 65.85 \\
 \cmidrule{2-6}
 & CNN & 89.48 & 86.66 & \textbf{88.81} & 87.17 \\ 
 \cmidrule{2-6}
 & LSTM & 88.50 & 85.09 & 87.35 & 85.64 \\ 
 \cmidrule{2-6}
 & ConvLSTM & \textbf{90.51} & \textbf{86.75} & 88.51 & \textbf{87.31} \\
\midrule 

%------------------------ 6 Activities ------------------------
\multirow{7}{*}{6} 
 & SVM & 76.34 & 80.11 & 82.02 & 77.75 \\ 
 \cmidrule{2-6}
 & MLP & 75.40 & 76.30 & 78.77 & 75.58 \\
 \cmidrule{2-6}
 & CNN & 92.60 & 91.59 & 93.67 & 91.83 \\ 
 \cmidrule{2-6}
 & LSTM & 94.03 & 92.87 & 95.19 & 93.55 \\ 
 \cmidrule{2-6}
 & ConvLSTM & \textbf{95.29} & \textbf{94.08} & \textbf{96.02} & \textbf{94.73} \\ 
\midrule 

%------------------------ 5 Activities ------------------------
\multirow{7}{*}{5} 
 & SVM & 76.63 & 83.76 & 81.19 & 79.86 \\ 
 \cmidrule{2-6}
 & MLP & 78.25 & 82.23 & 83.22 & 80.99 \\
 \cmidrule{2-6}
 & CNN & 92.57 & 94.09 & 94.40 & 93.79 \\ 
 \cmidrule{2-6}
 & LSTM & 93.99 & 95.08 & 95.08 & 94.79 \\ 
 \cmidrule{2-6}
 & ConvLSTM & \textbf{96.06} & \textbf{96.86} & \textbf{96.15} & \textbf{96.39} \\ 
\midrule 

%------------------------ 4 Activities ------------------------
\multirow{7}{*}{4} 
 & SVM & 88.80 & 92.16 & 88.53 & 89.08 \\ 
 \cmidrule{2-6}
 & MLP & 90.28 & 92.26 & 90.33 & 90.43 \\
 \cmidrule{2-6}
 & CNN & 96.28 & 97.12 & 96.17 & 96.49 \\ 
 \cmidrule{2-6}
 & LSTM & 96.91 & 97.61 & 96.79 & 97.11 \\ 
 \cmidrule{2-6}
 & ConvLSTM & \textbf{97.87} & \textbf{98.39} & \textbf{97.80} & \textbf{98.05} \\ 
\bottomrule
\end{tabular}
\end{table}

\begin{table}[!htbp]
\huge
\centering
\vspace{0.1cm}
\caption{Comparison of ConvLSTM model performance for 7 activity classification under the CSV approach.}
\label{model_performance_comparison1}
\resizebox{\linewidth}{!}{
\begin{tabular}{c|c|c|c|c|c|c|c}
\toprule
\multirow{2.3}{*}{\textbf{Metrics (\%)}} & \multicolumn{7}{c}{\textbf{Model Input}} \\ \cmidrule{2-8} 
 & \textbf{RD} & \textbf{RA} & \textbf{RE} & \textbf{RD+RA} & \textbf{RD+RE} & \textbf{RA+RE} & \textbf{RD+RA+RE} \\ \midrule
\textbf{Accuracy} & 76.94 & 71.76 & 89.53 & 86.26 & 90.40 & 87.51 & 90.51 \\ \cmidrule{1-8} 
\textbf{Precision} & 80.12 & 74.32 & 86.02 & 84.96 & 87.05 & 84.47 & 86.75 \\ \cmidrule{1-8} 
\textbf{Recall} & 79.98 & 77.18 & 87.33 & 87.16 & 88.95 & 86.42 & 88.51 \\ \cmidrule{1-8} 
\textbf{$F_1$-score} & 77.66 & 72.95 & 86.12 & 85.51 & 87.70 & 84.59 & 87.31 \\ \bottomrule
\end{tabular}
}
\end{table}

\begin{table}[!htbp]
\huge
\centering
\vspace{0.1cm}
\caption{Comparison of ConvLSTM model performance for 6 activity classification under the CSV approach.}
\label{model_performance_comparison2}
\resizebox{\linewidth}{!}{
\begin{tabular}{c|c|c|c|c|c|c|c}
\toprule
\multirow{2.3}{*}{\textbf{Metrics (\%)}} & \multicolumn{7}{c}{\textbf{Model Input}} \\ \cmidrule{2-8} 
 & \textbf{RD} & \textbf{RA} & \textbf{RE} & \textbf{RD+RA} & \textbf{RD+RE} & \textbf{RA+RE} & \textbf{RD+RA+RE} \\ \midrule
\textbf{Accuracy} & 82.70 & 78.22 & 94.09 & 92.60 & 95.58 & 91.23 & 95.29 \\ \cmidrule{1-8} 
\textbf{Precision} & 85.53 & 83.60 & 92.79 & 92.69 & 94.43 & 90.66 & 94.08 \\ \cmidrule{1-8} 
\textbf{Recall} & 87.80 & 86.96 & 95.14 & 94.74 & 96.43 & 93.67 & 96.02 \\ \cmidrule{1-8} 
\textbf{$F_1$-score} & 85.00 & 82.41 & 93.50 & 93.28 & 95.20 & 91.35 & 94.73 \\ \bottomrule
\end{tabular}
}
\end{table}

\begin{table}[!htbp]
\huge
\centering
\vspace{0.1cm}
\caption{Comparison of ConvLSTM model performance for 5 activity classification under the CSV approach.}
\label{model_performance_comparison3}
\resizebox{\linewidth}{!}{
\begin{tabular}{c|c|c|c|c|c|c|c}
\toprule
\multirow{2.3}{*}{\textbf{Metrics (\%)}} & \multicolumn{7}{c}{\textbf{Model Input}} \\ \cmidrule{2-8} 
 & \textbf{RD} & \textbf{RA} & \textbf{RE} & \textbf{RD+RA} & \textbf{RD+RE} & \textbf{RA+RE} & \textbf{RD+RA+RE} \\ \midrule
\textbf{Accuracy} & 78.42 & 79.59 & 95.93 & 88.37 & 95.48 & 93.28 & 96.06 \\ \cmidrule{1-8} 
\textbf{Precision} & 83.79 & 86.31 & 96.53 & 91.26 & 95.98 & 94.30 & 96.86 \\ \cmidrule{1-8} 
\textbf{Recall} & 82.43 & 87.10 & 96.11 & 92.28 & 95.73 & 95.11 & 96.15 \\ \cmidrule{1-8} 
\textbf{$F_1$-score} & 80.79 & 84.26 & 96.24 & 90.71 & 95.67 & 94.43 & 96.39 \\ \bottomrule
\end{tabular}
}
\end{table}

\begin{table}[!htbp]
\huge
\centering
\vspace{0.1cm}
\caption{Comparison of ConvLSTM model performance for 4 activity classification under the CSV approach.}
\label{model_performance_comparison4}
\resizebox{\linewidth}{!}{
\begin{tabular}{c|c|c|c|c|c|c|c}
\toprule
\multirow{2.3}{*}{\textbf{Metrics (\%)}} & \multicolumn{7}{c}{\textbf{Model Input}} \\ \cmidrule{2-8} 
 & \textbf{RD} & \textbf{RA} & \textbf{RE} & \textbf{RD+RA} & \textbf{RD+RE} & \textbf{RA+RE} & \textbf{RD+RA+RE} \\ \midrule
\textbf{Accuracy} & 86.49 & 95.43 & 96.81 & 96.49 & 97.45 & 97.66 & 97.87 \\ \cmidrule{1-8} 
\textbf{Precision} & 87.28 & 95.94 & 97.43 & 97.33 & 97.92 & 98.22 & 98.39 \\ \cmidrule{1-8} 
\textbf{Recall} & 86.89 & 95.53 & 96.67 & 96.30 & 97.37 & 97.49 & 97.80 \\ \cmidrule{1-8} 
\textbf{$F_1$-score} & 85.76 & 95.56 & 96.96 & 96.68 & 97.58 & 97.80 & 98.05 \\ \bottomrule
\end{tabular}
}
\end{table}

\subsubsection{LOPO-CV}
We report LOPO-CV results for ConvLSTM, which outperformed other models in the CSV approach. RD+RA+RE feature maps were used, as they generally yielded the best performance as well.
Table~\ref{leave_person_out} shows ConvLSTM’s performance in classifying activities using RD+RA+RE feature maps. Accuracy and $F_1$-score improve as the number of activities decreases, rising from $89.56\%$ ($87.15\%$) for 7 activities to $93\%$ ($91.12\%$) for 4 activities.  These results confirm the model’s subject-independent robustness, generalizability, and the effectiveness of RD+RA+RE maps.
Compared to the CSV approach, this method evaluates the model with a larger number of test samples. %Also, the performance analysis of the ConvLSTM model with single and pairwise combinations of feature maps was not conducted under this method.

\begin{table}[!htbp]
\huge
\centering
\vspace{0.1cm}
\caption{Comparison of ConvLSTM model performance presented as percentages using the LOPO-CV approach.}
\label{leave_person_out}
\resizebox{\columnwidth}{!}{%
\begin{tabular}{c|cc cc cc cc}
\toprule
 \multirow{2.3}{*}{\textbf{Test Subject}} & \multicolumn{2}{c}{\textbf{7 Activity}} & \multicolumn{2}{c}{\textbf{6 Activity}} & \multicolumn{2}{c}{\textbf{5 Activity}} & \multicolumn{2}{c}{\textbf{4 Activity}} \\ 
\cmidrule(lr){2-3}\cmidrule(lr){4-5}\cmidrule(lr){6-7}\cmidrule(lr){8-9}
 & \textbf{Accuracy} & \textbf{$F_1$-score} & \textbf{Accuracy} & \textbf{$F_1$-score} & \textbf{Accuracy} & \textbf{$F_1$-score} & \textbf{Accuracy} & \textbf{$F_1$-score} \\ 
\midrule
\textbf{Subject 1}   & 89.08    & 85.49    & 92.19    & 90.79    & 92.35    & 91.15    & 92.55    & 91.03 \\
\textbf{Subject 2}   & 93.14    & 91.23    & 96.90    & 96.25    & 97.96    & 97.89    & 97.95    & 97.49 \\
\textbf{Subject 3}   & 86.45    & 84.72    & 94.35    & 93.88    & 94.51    & 93.41    & 88.52    & 84.84 \\ 
\midrule
\textbf{Average}  & 89.56    & 87.15    & 94.48    & 93.64    & 94.94    & 94.15    & 93.00    & 91.12 \\ 
\bottomrule
\end{tabular}%
}
\end{table}

% To further demonstrate ConvLSTM's robustness with limited training data, we trained the model using RD+RA+RE from only one subject and evaluated it on a different, unseen participant for varying numbers of activities (Table~\ref{convlstm_performance_one_subject}). Despite the minimal training data, the model still performed well, demonstrating its ability to handle challenging scenarios with highly restricted datasets.

% \begin{table}[!htbp]
% \footnotesize
% \centering
% \vspace{0.1cm}
% \caption{Performance metrics for the ConvLSTM model trained on one subject with RD+RA+RE input and tested on an unseen participant across different activity sets.}
% \label{convlstm_performance_one_subject}
% \begin{tabular}{c|c|c|c|c}
% \toprule
% \multirow{2.3}{*}{\textbf{\#Activities}} 
%  & \multicolumn{4}{c}{\textbf{Metrics (\%)}} \\ 
% \cmidrule{2-5}
% & \textbf{Accuracy} & \textbf{Precision} & \textbf{Recall} & \textbf{\(F_1\)-score} \\ 
% \midrule
% % ------------------------ 7 Activities ------------------------
% 7 & 84.91 & 87.37 & 83.08 & 83.07 \\ 
% \midrule
% % ------------------------ 6 Activities ------------------------
% 6 & 94.46 & 94.16 & 95.21 & 94.54 \\ 
% \midrule
% % ------------------------ 5 Activities ------------------------
% 5 & 95.26 & 94.71 & 95.72 & 95.09 \\ 
% \midrule
% % ------------------------ 4 Activities ------------------------
% 4 & 93.35 & 92.75 & 94.03 & 93.15 \\ 
% \bottomrule
% \end{tabular}
% \end{table}

\section{Discussion}
\label{discussion}
The results of this study demonstrate the effectiveness of using FMCW radar for HAR through our proposed framework. In this framework, we utilized RD, RA, and RE feature maps across ML and DL models. The performance of the system is validated through two distinct strategies: CSV and LOPO-CV.

The ConvLSTM model outperformed other models, achieving an accuracy of $90.51\%$ and an \(F_1\)-score of $87.31\%$ on CSV, and an accuracy of $89.56\%$ and an \(F_1\)-score of $87.15\%$ on LOPO-CV. ConvLSTM combines CNN and LSTM strengths by integrating convolution into LSTM gates, capturing spatiotemporal dependencies from RD, RA, and RE data. It excels in distinguishing similar activities where both spatial and motion cues are crucial, such as “sitting on a bed” versus “sitting on a chair.” The steady decrease in training and validation loss, as illustrated in Fig.~\ref{convlstm_ls}, further confirms the model's generalization ability and resistance to overfitting. These findings highlight the potential of the proposed approach for scalable, non-intrusive, and privacy-preserving activity monitoring in real-world scenarios.

One of the main contributions of this study is the introduction of a novel approach that directly feeds multi-dimensional radar feature maps into the model. Unlike conventional approaches that process feature maps as images, our method preserves the spatial and temporal structures of the data by treating RD, RA, and RE feature maps as data vectors. This approach successfully develops a robust HAR model using data from only three subjects, making it particularly suitable for real-world applications where large-scale data collection is often infeasible.

Our findings advance the current state of knowledge in the field by demonstrating that the proposed solution can achieve strong performance even when trained on a limited dataset. This is especially valuable for applications in healthcare monitoring and smart environments, where non-intrusive activity monitoring is essential. The practical implications of our work include the potential for real-time activity recognition and monitoring in various domains, such as elderly care, rehabilitation, and smart home automation.

% Notably, the RE feature map outperformed RD and RA, emphasizing the importance of vertical spatial information. Pairwise combinations like RD+RE enhanced performance, except in 5-activity classification, highlighting the complementarity of motion and spatial features. Adding RA to RD+RE yielded minimal gains, suggesting its contribution may be redundant.

% In contrast to previous studies that used image processing techniques on feature maps, this study introduces a novel approach by directly feeding extracted RD, RA, and RE feature maps into the model. This method successfully develops a robust HAR model using data from only three subjects, demonstrating the efficiency of the data processing methodology and the ConvLSTM architecture's ability to capture intricate spatiotemporal patterns. As a result, this approach is particularly suitable for real-world applications, including healthcare monitoring and smart environments, where large-scale data collection is often infeasible. The strong performance of ConvLSTM under LOPO-CV further supports these findings.

\section{Conclusion}
\label{conclusion}

In this paper, we proposed an FMCW radar-based framework for HAR in home-like environments. The study demonstrated the feasibility and advantages of integrating multi-dimensional feature maps as data vectors to effectively capture the temporal and spatial characteristics of human activities. By leveraging the advanced capabilities of FMCW radar systems in conjunction with DL models, particularly the ConvLSTM architecture, we successfully extracted spatiotemporal patterns from radar feature maps, thereby surpassing traditional ML and DL approaches in performance. 

Our results emphasize the benefits of using multiple feature maps for improved recognition accuracy, with RE proving the most effective, highlighting the role of vertical spatial information in precise activity recognition. Furthermore, the ConvLSTM model effectively captured both motion and spatial dependencies, demonstrating robust performance across diverse and challenging scenarios and when tested on unseen data. The proposed solution demonstrated strong performance even when trained on a limited dataset, making it suitable for real-world applications in domains such as healthcare and smart homes that require non-intrusive activity monitoring. This advantage is particularly valuable when comprehensive data collection may be impractical due to domain-specific constraints. 

Despite promising results, the proposed approach has certain limitations. Therefore, future research should focus on addressing these limitations such as the limited dataset diversity and a restricted range of activities. Additionally, the computational cost of these models has not been calculated, though it is a crucial metric for large-scale implementation, real-time applications, and deployment on edge devices. Addressing these limitations is essential to enhance the robustness and applicability of the proposed framework in real-world scenarios.

In conclusion, this study presents a novel FMCW radar-based framework for HAR that leverages multi-dimensional feature maps and advanced DL models. The proposed approach demonstrates strong performance and practical applicability, paving the way for future research and development in the field of non-intrusive activity monitoring.

\section*{Acknowledgments}
The authors would like to thank all participants who participated in this study. The authors also thank Mälardalen University, the Schlegel-UW Research Institute for Aging, Gold Sentintel Inc, NSERC, and MITACS for supporting this study.

\bibliographystyle{IEEEtran}
\bibliography{main}

\end{document}